\documentclass[%
 aps,
 amsmath,amssymb,
superscriptaddress,
 reprint,%
]{revtex4-2}

\usepackage{comment}
\usepackage{graphicx}
\usepackage{dcolumn}
\usepackage{bm}

\usepackage{xcolor}

\newcolumntype{L}[1]{>{\raggedright\arraybackslash}m{#1}}
\newcolumntype{C}[1]{>{\centering\arraybackslash}m{#1}}
\newcolumntype{R}[1]{>{\raggedleft\arraybackslash}m{#1}}

\usepackage{sistyle}
\SIthousandsep{,}

\begin{document}

\preprint{AIP/123-QED}

\title{Laser Wakefield Acceleration Driven by a Discrete Flying Focus}

\author{Jacob R. Pierce}
    \email{jacobpierce@physics.ucla.edu}
    \affiliation{Department of Physics and Astronomy, University of California, Los Angeles, California 90095, USA}

\author{Kyle G. Miller}
    \affiliation{University of Rochester, Laboratory for Laser Energetics, Rochester, New York 14623, USA}
    
\author{Fei Li}
    \affiliation{Department of Engineering Physics, Tsinghua University, Beijing 100084, China}
 
\author{John P. Palastro}
    \email{jpal@lle.rochester.edu}
    \affiliation{University of Rochester, Laboratory for Laser Energetics, Rochester, New York 14623, USA}

\author{Warren B. Mori}
    \affiliation{Department of Physics and Astronomy, University of California, Los Angeles, California 90095, USA}

\date{\today}

\begin{abstract}

Laser wakefield acceleration (LWFA) may enable the next generation of TeV-scale lepton colliders. Reaching such energies will likely require multiple LWFA stages to overcome limitations on the energy gain achievable in a single stage. The use of stages, however, introduces challenges such as alignment, adiabatic matching between stages, and a lower average accelerating gradient.
Here, we propose a discrete flying focus that can deliver higher energy gain in a single stage, thereby reducing the number of stages required for a target energy. A sequence of laser pulses with staggered focal points and delays drives a plasma wave in which an electron beam experiences a near-constant accelerating gradient over distances beyond those attainable with a  conventional pulse. Simulations demonstrate that a discrete flying focus with a total energy of 150~J can  transfer 40~GeV per electron to a 50-pC beam in a single 30-cm stage, corresponding to 50~dephasing lengths.

\end{abstract}

\maketitle

\section{Introduction}
\label{sec:introduction}

\begin{figure}[t]
        \centering
        \includegraphics[width=\linewidth]{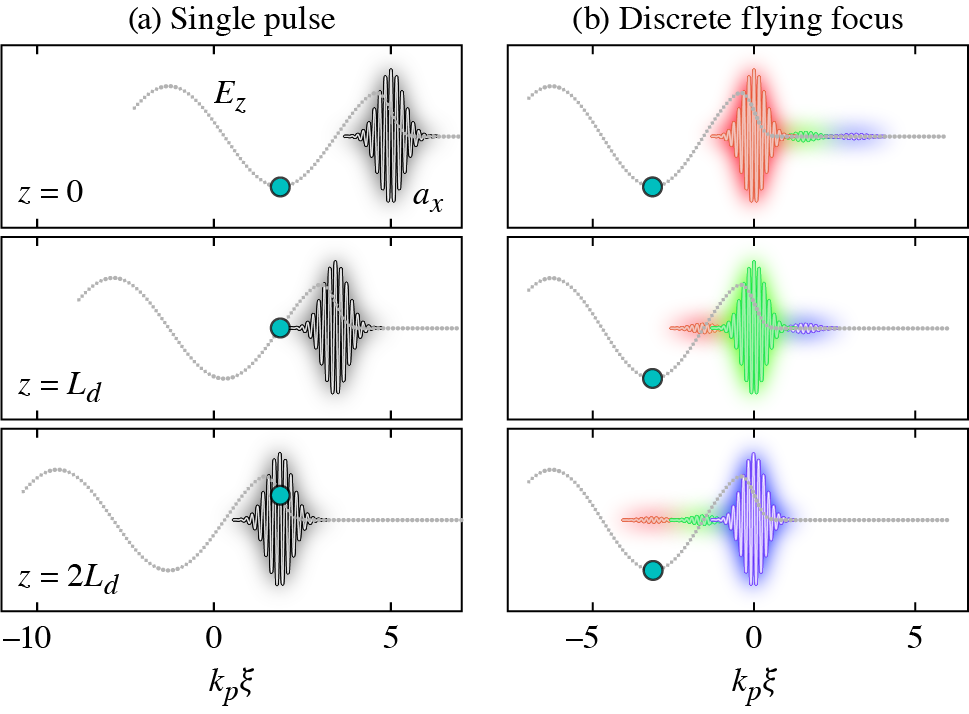}
        \caption{Comparison of LWFA with a conventional pulse and a discrete flying focus in the linear regime. (a) A single pulse traveling at its group velocity (black) drives a plasma wave (gray), whose longitudinal field $E_z$ accelerates a highly relativistic electron (cyan). The pulse and longitudinal field slip backward in $\xi = z - ct$ as the pulse propagates from $z = 0$ to $2L_d$. As a result, the electron moves from an accelerating to a decelerating longitudinal field, which limits its energy gain---a process referred to as dephasing. (b) A sequence of pulses (red, green, blue) with appropriately selected delays and focal points drives a plasma wave (gray) whose longitudinal field remains stationary in $\xi$. This eliminates dephasing by ensuring that the electron (cyan) remains collocated with the maximum accelerating field over the entire distance.
        In both (a) and (b), the field of the laser pulse is calculated using Eq.~\eqref{eq:astrl}, and the longitudinal field is calculated using linear theory \cite{gorbunov1987excitation,esarey2009physics}. The specific parameters, chosen for illustration, are $k_0/k_p =40$, $k_p \tau_0 = 0.5$, $f_j = jL_d$, and $\Delta_j = (1-v_g/c) f_j = \left(\frac \pi 2\right)j k_p^{-1} $. The dephasing length is $k_p L_d = 1600\pi$, the single pulse in (a) has $k_p z_R = \infty$, and each pulse in (b) has $k_p z_R = 320$.} 
        \label{fig:schematic}
 \end{figure}

    In laser wakefield acceleration (LWFA), a laser-driven plasma wave focuses and accelerates an electron beam \cite{tajima1979laser}. Because plasma waves can sustain accelerating gradients that are orders of magnitude higher than those of conventional radio-frequency accelerators, LWFA has been considered for a next-generation TeV-scale lepton collider \cite{schroeder2010physics,schroeder2023linear}. However, the required TeV-scale beam energies are far higher than the GeV-scale energies that can be transferred to a beam in a single LWFA stage. This is due to two limitations: (i) dephasing [Fig. \ref{fig:schematic}(a)], where the beam outruns the accelerating phase of the plasma wave, and (ii) pump depletion, where the laser pulse exhausts its energy by driving plasma waves. To overcome these limitations, the advanced accelerator community envisions chaining together multiple LWFA stages. For example, the Snowmass 2021 Design Frontier proposed an electron accelerator with an accelerating gradient of ${\sim}100 \text{ GeV/m}$, but only ${\sim} 3 \text{ GeV}$ of energy gain per 3-cm stage \cite{schroeder2023linear}. With these parameters, ${\sim} 300$ stages would be required to accelerate an electron beam to $1 \text{ TeV}$. 
    
    The use of stages for LWFA poses significant challenges, including beam transport between stages, adiabatic matching of the beam into each plasma section, and alignment and temporal synchronization of the beam and laser pulses \cite{esarey2009physics,mehrling2012transverse,steinke2016multistage,zhao2020emittance}. In addition, the meter-scale gaps required between each stage would reduce the average accelerating gradient by nearly two orders of magnitude. These challenges, along with the overall collider complexity, can be mitigated by decreasing the number of stages $N_\text{stages}$. One approach is to increase the energy gain in each stage by operating at lower electron densities: $\Delta E_\text{stage} \propto \smash{n_0^{-1}}$ so that $N_\text{stages} \propto n_0$. However, this approach decreases the accelerating gradient $\smash{(E_z \propto \sqrt{n_0})}$ and hence increases the total length of plasma $L_T \propto \smash{n_0^{-1/2}}$\cite{esarey2009physics}. 

    Space-time structured laser pulses offer an alternative approach to enhancing the energy gain in a single LWFA stage \cite{debus2019circumventing,palastro2020dephasingless,caizergues2020phase,miller2023dephasingless,nutting2025transversely,liberman2025}. These ``flying focus'' concepts rely on a controllable-velocity intensity peak to keep the accelerating phase of the driven plasma wave collocated with an electron beam, thereby eliminating dephasing. This allows for operation at higher densities, enabling larger energy gains over the same length or identical energy gains over shorter lengths when compared to equal-energy conventional pulses. By providing higher energy gains, such concepts could reduce the number of required stages without increasing the length of the plasma. Flying-focus pulses have been experimentally demonstrated using two techniques:  (i) chromatic focusing of a chirped laser pulse \cite{froula2018spatiotemporal,Jolly2020} and (ii) reflection from an axiparabola--echelon pair, which independently focuses and delays radial annuli of a pulse \cite{Pigeon2024}. This latter technique is now being considered as the basis for an experiment to achieve a 100-GeV-scale electron beam in a single meter-scale stage as part of the NSF OPAL project \cite{NSF_Award_2329970,shaw2025}. 
    
Here, we introduce a concept for enhancing the energy gain in a single LWFA stage using a train of collinear laser pulses with different focal points [Fig. \ref{fig:schematic}(b)]. The delays between the pulses in this discrete flying focus (DFF) are chosen so that each pulse comes into focus and drives the plasma wave a fixed distance ahead of the electron beam. Despite the subluminal group velocity of each pulse, the accelerating phase of the plasma wave advances at the vacuum speed of light. As a result, the accelerating phase remains collocated with a highly relativistic electron beam over the entire stage. In the linear regime (normalized vector potentials $a_0 \lesssim 0.5)$, each pulse drives the plasma wave strongly over its Rayleigh range. In the nonlinear regime ($a_0 \gtrsim 2)$, each pulse self-guides and strongly drives the plasma wave over its pump depletion length. Quasistatic particle-in-cell simulations in this regime show that a DFF can drive a plasma wave that accelerates a {50-pC} electron beam by {40 GeV} in a single {30-cm} stage, corresponding to 50 dephasing lengths.

The remainder of this article begins with theory describing propagation of a DFF pulse (Sec. II). The theory extends the Arbitrarily Structured Laser (ASTRL) pulse concept \cite{pierce2023arbitrarily} to account for plasma dispersion. The pulse and plasma parameters for an LWFA stage far longer than a dephasing length are designed using scaling laws in the linear and nonlinear regimes (Sec. II). QPAD simulations \cite{li2021quasi,li2022integrating} based on these designs confirm that a plasma wave driven by a DFF can accelerate an externally injected electron beam over tens of dephasing lengths in both regimes (Sec. III). Notably, the simulations are not extensively optimized, indicating that further tuning of the DFF parameters could result in even larger energy gains or higher beam charge. Further, the ability to independently tune the parameters of each pulse (e.g., polarization, spot size, and duration) offers greater flexibility to structure and control the plasma wave than previous flying-focus concepts \cite{debus2019circumventing,palastro2020dephasingless,caizergues2020phase,miller2023dephasingless,liberman2025}. These prospects are discussed in the concluding section of the paper (Sec. IV).

\section{Theory}
\label{sec:theory}

    The design of a DFF for LWFA begins with an extension of the formalism developed for ASTRL pulses ~\cite{pierce2023arbitrarily} to account for linear dispersion in cold, uniform plasma.
    The propagation of a laser pulse through such a plasma is described by the wave equation
    \begin{equation}
        \left[ \nabla^2 - \frac 1 {c^2} \partial_t^2 \right] \mathbf A = \frac{ \omega_p^2}{c^2} \mathbf A,
        \label{eq:linear-dispersion}
    \end{equation}
    where $\mathbf A$ is the vector potential of the pulse in the Coloumb gauge, $\nabla \cdot \mathbf{A} = 0$, and $\omega_p = (e^2n_0/m_e\varepsilon_0)^{1/2}$ is the plasma frequency. For propagation along the $\hat{\boldsymbol z}$ direction, the transverse vector potential can be expressed as a normalized vector envelope $\mathbf a( t, z, \mathbf x_\perp )$ modulated by a carrier wave
     \begin{equation}
             \mathbf A_\perp( t, z, \mathbf x_\perp ) = \tfrac 1 2 ( \tfrac{m_ec^2}{e} )  {\mathbf a}( t, z, \mathbf x_\perp ) e^{ik_0 (z- v_\phi t)} + \text{c.c.},
        \label{eq:envelope_def}
        \end{equation}
    where $k_0$ is the central wavenumber and $v_\phi = \omega_0 / k_0 $ is the phase velocity, with $\omega_0^2 = \omega_p^2 + c^2 k_0^2$. 
    Because the transverse components of the vector potential evolve independently, it is sufficient to consider the evolution of a single component: $\mathbf a(t,z,\mathbf x_\perp) = a( t,z,\mathbf x_\perp) \hat{ \boldsymbol \epsilon } $ where $\hat{\boldsymbol \epsilon}$ is a polarization unit vector. Substituting Eq.~\eqref{eq:envelope_def} into Eq.~\eqref{eq:linear-dispersion} and transforming to the coordinates $(\zeta,s) \equiv (z - v_g t, z)$, where $v_g = c^2 k_0 / \omega_0 \approx \left[1 - \frac 1 2 ( \omega_0 / \omega_p )^2 \right]c $ is the linear group velocity, yields 
    \begin{equation}
        \begin{split}
            \Big[ \partial_s^2 + 2 \partial_s  \partial_\zeta &+ \left( 1 - \frac{v_g}{c} \right)^2 \partial_\zeta^2  
            \\ & + \nabla_\perp^2 + 2ik_0 \partial_s  \Big] a( \zeta, s, \mathbf x_\perp ) = 0.
        \end{split}
        \label{eq:env-pde-full}
    \end{equation} 
    The first three terms in Eq. \eqref{eq:env-pde-full} are dropped since they are small compared to the $\nabla_\perp^2 $ and $2ik_0 \partial_s$ terms (see Appendix~\ref{sec:higher-order-effects}).  This reduces Eq.~\eqref{eq:env-pde-full} to the steady-state paraxial wave equation
    \begin{equation}
        \left[ \nabla_\perp^2 + 2 i k_0 \partial_s  \right] a( \zeta, s, \mathbf x_\perp) \approx 0.
        \label{eq:env-pde-approx}
    \end{equation}
    The absence of $\partial_\zeta$ terms in
     Eq.~\eqref{eq:env-pde-approx} 
    implies that each temporal slice of the laser pulse evolves independently. Thus, Eq.~\eqref{eq:env-pde-approx} admits separable solutions of the form  
    \begin{equation}
        a( t, z, \mathbf x_\perp ) = \mathcal{A}_0 B( z - v_g t ) C( z, \mathbf x_\perp ),
        \label{eq:singlepulse}
    \end{equation}
    where the coordinates $(t,z)$ have been reintroduced. Here, $\mathcal{A}_0$ is the amplitude, $B$ is an arbitrary complex function, and $C$ is a complex function satisfying the paraxial wave equation $[ \nabla_\perp^2 + 2 ik_0 \partial_z ] C = 0$. This is the same form derived for vacuum propagation in Ref.~\cite{pierce2023arbitrarily} with $v_g$ replacing the vacuum speed of light in $B$.
    
    A laser pulse with an envelope described by Eq.~\eqref{eq:singlepulse} will excite a plasma wave with a relativistic phase velocity when propagating in an underdense plasma ($\omega_p \ll \omega_0$).
    In the linear regime $(a_0 \equiv \mathrm{max}|a| \lesssim 0.5 )$, 
    the distance over which a plasma wave excited 
    by the solution in Eq.~\eqref{eq:singlepulse} can accelerate a particle with $v_z \approx c$
    is limited by diffraction and 
    dephasing. Diffraction is associated with the transverse profile $C(z,\mathbf x_\perp)$, which determines the length scale over which the pulse maintains a high intensity and strongly drives the plasma wave. Dephasing is associated with the temporal profile $B(z-v_gt)$, which determines the phase velocity of the plasma wave  $v_w = v_g$. The wavelength of the wave is $ \lambda_w = 2\pi k_p^{-1}$, where $k_p \equiv \omega_p / c$ is the plasma wavenumber. Thus, electrons with $v_z \approx c$ initially collocated with the maximum accelerating field of the wave will encounter a decelerating field when the laser pulse has slipped backward by $ \Delta \xi = (\pi / 2 ) k_p^{-1} $, where $\xi \equiv z - ct$. This occurs after a distance referred to as the dephasing length $L_d$, given by
    \begin{equation}
        k_p L_d \equiv \frac \pi 2 \left( 1 - \frac{v_w}{c} \right)^{-1}.
        \label{eq:Ld}
    \end{equation}
    For $\omega_0 / \omega_p \gg 1$, $ k_p L_d \approx \pi (\omega_0 / \omega_p)^2 $. Figure ~\ref{fig:schematic}(a) illustrates dephasing in the linear regime for a single pulse with $C=1$. The pulse (black) and accelerating field of the plasma wave ($E_z$, gray dotted line) slip backward with respect to the electron (cyan) such that it is no longer accelerated for $z > L_d$. 
    
    The key idea of the present work is that dephasing can be overcome by using a train of collinear pulses with different focal points. To demonstrate this, a linear superposition of $N$ laser pulses of the form given by Eq.~\eqref{eq:singlepulse} is considered:
    \begin{equation}
        \begin{split}
            \mathbf a(t, z, \mathbf x_\perp ) \approx 
            \sum_{j=1}^N \tfrac 1 2 \mathcal{A}_{0j} B_j( &z - v_g t) C_j( z, \mathbf x_\perp ) 
            \\ & \times e^{ ik_0(z-v_\phi t)} \hat{ \boldsymbol \epsilon }_j + \text{c.c.} .
            \label{eq:astrl}
        \end{split}
    \end{equation}
    The amplitude of each pulse $\mathcal{A}_{0j}$ is chosen to produce a desired amplitude $a_0$ for the overall structure.
    The profiles $B_j$ and $C_j$ are selected so that each pulse comes into focus when it has slipped backward into the value of $\xi$ where the previous pulse was focused. Here, as in Ref.~\cite{pierce2023arbitrarily}, a Gaussian temporal profile is used:  
    \begin{equation}
        B_j(z - v_gt ) = \exp \left[ - \left(\frac{z - v_g t - \Delta_j }{ c\tau_j } \right) ^2 \right],
        \label{eq:longitudinal-profile}
    \end{equation}
    where $\tau_j$ is the duration and $\Delta_j$ is the initial peak position of each pulse. 
    Similarly, the paraxial solution for a Gaussian beam is used for the transverse profile:
    \begin{equation}
        C_j(z, \mathbf x_\perp) = \left( \frac{z_{R,j}}{q_j(z)} \right) \exp \left[ - \frac{ik_0 |\mathbf x_\perp  |^2 }{2q_j(z)} \right],
        \label{eq:gaussian-beam}
    \end{equation}
    where $q_j(z) \equiv z - f_j + i z_{R,j} $ is the complex beam parameter, $f_j$ is the focal point, $z_{R,j} = \frac 1 2 k_0 w_{0j}^2 $ is the Rayleigh range, and $w_{0j}$ is the spot size of each pulse. For simplicity, the same amplitude $\mathcal{A}_{0j} = \mathcal{A}_0$, Rayleigh range $z_{R,j} = z_R$, and duration $\tau_{j} = \tau_0$ are used for each pulse. 

    With these choices, the overall amplitude of the envelope (Eq.~\eqref{eq:astrl}) peaks when each pulse comes into focus at $z_{\text{focus},j}=f_j$ and $t_{\text{focus},j} = (f_j - \Delta_j ) / v_g$. The corresponding $\xi$ value of the peak amplitude is then given by $\xi_{\text{focus},j} = z_{\text{focus},j} - c t_{\text{focus},j} = (c/v_g) \Delta_j + (1 - c / v_g)f_j$. Thus, each pulse will come into focus at the same value of  $\xi_{\text{focus},j} = \xi_0$ if the $\Delta_j$ are chosen as
    \begin{equation}
            \Delta_j = \left( \frac{v_g}{c} \right) \xi_0 
            + \left( 1 - \frac{v_g}{c} \right) f_j.
        \label{eq:delta-j-linear}
    \end{equation}
    In the limit of many pulses ($N\gg1$), this choice for $\Delta_j$ produces a continuous peak that moves through the plasma at the vacuum speed of light $v_f =c$ (or through vacuum at the superluminal velocity $v_f = c^2 / v_g$). In practice, the phase velocity of the plasma wave $v_w$ may be slightly different from $v_f$ due to nonlinear effects, linear effects absent in Eq.~\eqref{eq:astrl} (e.g., higher-order dispersion or space--time couplings), and the dependence of the plasma wave structure on the entire pulse profile rather than its peak location. Nevertheless, Eq.~\eqref{eq:delta-j-linear} provides a starting point that can be tailored to compensate for these effects and achieve $v_w = c$.

    Figure~\ref{fig:schematic}(b) illustrates how dephasing can be circumvented in the linear regime using a DFF with $N=3$ pulses. The schematic shows the transverse vector potential of the three pulses [Eq.~\eqref{eq:astrl}] and the longitudinal electric field $E_z$ of the plasma wave that they drive. As each pulse slips backward in $\xi$, it is replaced by its successor, which becomes the predominant driver of the plasma wave. With $\Delta_j$ and $f_j$ related by Eq.~\eqref{eq:delta-j-linear}, the $\xi$ value of the peak accelerating field remains stationary and collocated with a highly relativistic electron (cyan) over multiple dephasing lengths ($z=0$ to $2L_d$).  
    
    The total length that the DFF drives the plasma wave $L_\mathrm{D}$ is determined by the number of pulses $N$, the distance between successive focal points $\delta f_j = f_{j+1} - f_j$, and the distance over which a single pulse drives the plasma wave $L_p$. For equally spaced focal points, $\delta f_j = \delta f$ and 
    \begin{equation}
    L_\mathrm{D} = (N-1) \delta f + L_p.  
    \end{equation}\label{eq:LD}
    A minimum number of pulses $N_\text{min}$ is required to maintain a near-uniform accelerating field over the length $L_\mathrm{D}$. This minimum can be estimated from the condition that the distance between successive focal points is less than the distance over which a single pulse drives the plasma wave: $\delta f < L_p$. A single pulse maintains a high amplitude and drives the plasma wave over a Rayleigh range $z_R$ in the linear regime ($a_0 \lesssim 0.5$) and a pump depletion length $L_\text{pd}$ in the nonlinear regime ($a_0 \gtrsim 2$) \cite{lu2007generating}. For a nonlinearly matched pulse with $k_p w_0 = 2 \sqrt{a_0}$, the pump depletion length is given by ${L_\text{pd} = (k_0 / k_p)^2 c \tau_0 }$ \cite{lu2007generating}.  Thus, the minimum number of pulses required for near-uniform acceleration may be estimated as 
    \begin{equation}
        N_\text{min} 
        \approx \begin{cases}
            L_\mathrm{D}/z_R, \quad & a_0 \lesssim 0.5
            \\ L_\mathrm{D}/L_\text{pd}, \quad & a_0 \gtrsim 2
        \end{cases}.
        \label{eq:Nmin}
    \end{equation}
    In most cases of relevance, $z_R \ll L_\mathrm{pd}$, such that a DFF would require many more pulses in the linear regime than the nonlinear regime.

    A highly relativistic electron ($v_z \approx c$) that experiences a nearly uniform accelerating field over a distance $L$ will gain an energy $m_e c^2 \Delta \gamma \approx -eE_z L$. In the linear and nonlinear regimes,
    \begin{equation}
        \Delta \gamma 
        \approx \begin{cases}
            \tfrac{\pi}{8}k_pLa_0^2, \quad & a_0 \lesssim 0.5
            \\ \tfrac{1}{2}k_pL\sqrt{a_0}, \quad & a_0\gtrsim 2
        \end{cases}.
        \label{eq:egain}
    \end{equation}
    With a conventional pulse, $L$ is limited to a dephasing length: either $k_pL_d = \pi (k_0/k_p)^2$ in the linear regime or $k_pL_d = \tfrac{4}{3}(k_0/k_p)^2\sqrt{a_0}$ in the nonlinear regime \cite{lu2007generating}. While the dephasing length can be increased by using a density gradient \cite{katsouleas1986}, the acceleration length would still be limited by pump depletion, which also scales as $k_0^2/k_p^2$. Thus, for fixed $a_0$ and $k_0$ in the conventional case, increasing the energy gain in a single stage, $\Delta \gamma_\mathrm{C} \propto \smash{n_0^{-1}}$, requires operating at lower electron densities with weaker accelerating fields, $E_z \propto \smash{\sqrt{n_0}}$, and longer stages, $L_d \propto \smash{n_0^{-3/2}}$.
    
    With the DFF, $L = L_\mathrm{D} \approx N \delta f$, which is limited only by the number of pulses when $\delta f$ is fixed. As a result, the singe-stage energy gain can be increased without resorting to lower densities and weaker fields by using more pulses. For $\delta f = z_R$ in the linear regime and $\delta f= L_\mathrm{pd}$ in the nonlinear regime, 
    \begin{equation}
        \Delta \gamma_{\mathrm D} 
        \approx \begin{cases}
            \dfrac{\pi}{8}k_pz_RNa_0^2, \quad & a_0 \lesssim 0.5
            \\ \dfrac{k_0^2}{2k_p} c\tau_0 N \sqrt{a_0}, \quad & a_0\gtrsim 2
        \end{cases}.
        \label{eq:egainDFF}
    \end{equation}
    In this configuration, the energy gain increases linearly with the number of pulses $N$ and has a more favorable scaling with the density compared to a conventional pulse: $\Delta \gamma_\mathrm D \propto \sqrt{n_0}$ and $\propto 1/\sqrt{n_0}$ in the linear and nonlinear regimes, respectively. 
    
    By eliminating dephasing, LWFA driven by a DFF can operate at higher densities where the accelerating field is stronger: $E_z \propto \smash{\sqrt{n_0}}$. This allows for (i) larger energy gains over the same length, (ii) equal energy gains over shorter lengths, or (iii) anything in between. Option (i) would minimize the number of stages required to reach a target energy. In a single stage of fixed length $L$, the ratio of energy gains attainable by a DFF and conventional pulse with the same $a_0$ and $k_0$ is given by $\Delta \gamma_\mathrm{D}/\Delta \gamma_\mathrm{C}=k_{pD}/k_p$ where $k_{p\mathrm D} = (e^2 n_{0\mathrm D}/m_e \varepsilon_0c^2)^{1/2}$ and $n_{0\mathrm D}$ is the design density for LWFA driven by a DFF, which can be much higher than in LWFA with a conventional pulse\cite{palastro2020dephasingless}. For a conventional pulse, the value of $k_p$ can be related to the acceleration (dephasing) length, leading to the useful expression \cite{palastro2020dephasingless}:
    \begin{equation}
        \frac{\Delta \gamma_{\mathrm D}}{\Delta \gamma_{\mathrm C}} 
        \approx \left(\dfrac{k_{p\mathrm D}}{k_0}\right)
        \begin{cases}
             \left(\dfrac{k_{0} L}{\pi}\right)^{1/3} & a_0 \lesssim 0.5
            \\  \left(\dfrac{3k_{0}L}{4\sqrt{a_0}}\right)^{1/3}  & a_0\gtrsim 2
        \end{cases}.
        \label{eq:egainRatio}
    \end{equation}
    
    As an example, in a single stage with $L=32$ cm, $n_{0\mathrm D} = 3\times10^{18}$ $\mathrm{cm}^{-3}$, $a_0 = 4$, and $\lambda = 2\pi c/\omega_0 = 800$ nm, $\Delta \gamma_{\mathrm D} \approx 4\Delta \gamma_{\mathrm C}$---the energy gain attainable with a DFF is approximately 4$\times$ higher than that attainable with a conventional pulse. The advantage of the DFF would be even larger at higher densities or with longer stages, albeit at the cost of more pulses. We note that while the ratio of energy gains increases as the density increases, the amount of charge that can be loaded into the wakefield decreases as $n_{0D}^{-1/2}$, so there is a balance between acceleration gradient and the amount of charge that can be loaded into the wakefield \cite{lu2007generating, tzoufras2008beam,dalichaouch2021}.
    
    The number of pulses composing a DFF $N$ 
    also controls the amount of overlap between pulses (for fixed $\Delta_j$ and $\tau_j$). When the pulses overlap significantly, the profile of the overall envelope [Eq. \eqref{eq:astrl}] is sensitive to the relative phases of the pulses and may be different than that of the individual pulses. For example, the effective spot size $w_\text{eff}$ may be larger than the spot size of each pulse $w_0$. The pulses begin to overlap when their spacing is comparable to the pulse duration, i.e., $\Delta_{j+1} - \Delta_j \approx c\tau_0$. The number of pulses beyond which overlapping occurs may be estimated as
    \begin{equation}
        N_\text{overlap} = \left( 1 - \frac{v_g}{c} \right) \frac{L}{c\tau_0}.
        \label{Eq:Noverlap}
    \end{equation}
    In the nonlinear regime, $N_\text{overlap} \approx \frac 1 2 N_\text{min}$. This shows that the pulses at least partially overlap in the nonlinear regime if the condition $N>N_\mathrm{min}$ is satisfied, regardless of the pulse or plasma parameters. In the linear regime, $N_\text{overlap} \approx \tfrac{1}{4}(k_p^2 w_0^2 / \omega_0 \tau_0 ) N_\text{min}$. Here, the degree of overlap depends on the pulse and plasma parameters. In either case, overlap can be mitigated by choosing the polarization of successive pulses to be orthogonal, which effectively doubles $N_\text{overlap}$.

        \begin{table}[t]
            \renewcommand{\arraystretch}{1.1}
            \begin{tabular}{
            L{0.4\linewidth} L{0.3\linewidth} L{0.3\linewidth} }
                \hline 
                Laser parameters & Value & Normalized \\
                \hline
                $\lambda_0$ & 800 nm & $2\pi/25$ \\
                $k_0$ & 7.8 $\mu\text{m}^{-1}$ & $25$ \\
                $\tau_0$ & 16 fs & $1.5$ \\
                $w_0$ & 13 $\mu$m & $4$ \\
                $z_R$ & 640 $\mu$m & 200 \\
                $L_d $ & 0.64 cm & $\num{2000}$ \\
                $L_\text{pd}$ & 0.32 cm & $\num{1000}$ \\ 
                \hline
                Discrete flying focus parameters ($a_0 = 0.1$) \\ 
                \hline
                $ N $ & -- & 40, 250 \\
                $ N_\text{min}$  & -- & 250 \\ 
                $ N_\text{overlap}$ & -- & 25 \\
                $a_{0j}$ & -- & 0.1, 0.024 \\
                $f_{j+1} - f_j$ & 4 mm, 0.6 mm & 1250, 200 \\
                $\Delta_{j+1} - \Delta_j $ & 3.2 $\mu$m, 0.5 $\mu$m & $1, \; 0.16$ \\
                \hline
                Discrete flying focus parameters ($a_0=4$) \\ 
                \hline
                $ N $ & -- & 20, 40 \\
                $ N_\text{min} $ & -- & 50 \\
                $ N_\text{overlap}$ & -- & 25 \\
                $a_{0j}$ & -- & 4 \\
                $f_{j+1} - f_j$ & 8 mm, 4 mm & 2500, 1250 \\
                $\Delta_{j+1} - \Delta_j $ & 6.4 $\mu$m, 3.2 $\mu$m & 2, 1 \\
                \hline
                Plasma parameters & &  \\
                \hline
                $n_0$ & $3\times10^{18} \text{ cm}^{-3}$ & -- \\
                $k_p^{-1}$ & 3.2 $\mu$m & 1 \\ 
                $L$ & 16 cm & $ \num{50000} $  \\
                \hline 
            \end{tabular}
        \caption{\label{tab:ez-onaxis-params} Parameters for the simulations presented in Figs.~\ref{fig:Eenv-Ez-snapshots} and \ref{fig:ez-onaxis}. The normalized vector potential of each pulse $a_{0j}$ is scaled to achieve the desired $a_0$ for the overall envelope. Note that for the profile choices defined in Eqs.~\eqref{eq:longitudinal-profile} and ~\eqref{eq:gaussian-beam}, $a_{0j} = \mathcal{A}_{0j}$. The values for $f_{j+1} - f_j$ and $\Delta_{j+1} - \Delta_j$ are respectively listed for the $N>1$ cases. Time and space are normalized by $\omega_p^{-1}$ and $k_p^{-1}$ where indicated.}
    \end{table}    
    
\begin{figure*}[t]
        \centering
        \includegraphics[width=\linewidth]{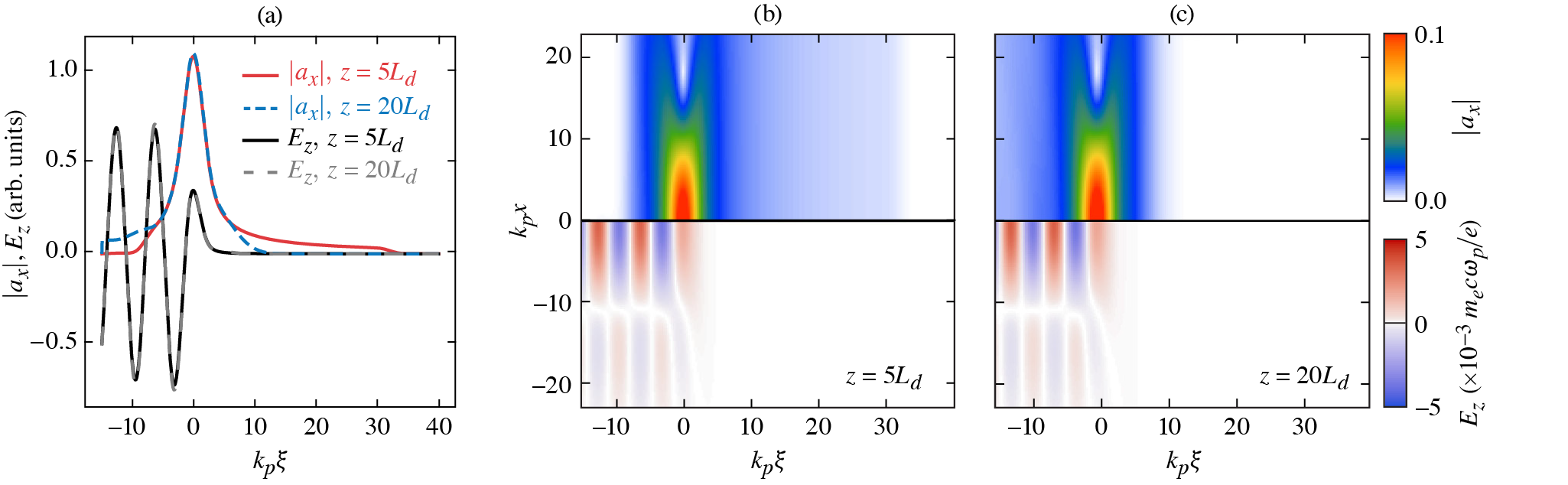}
        \caption{Evolution of a discrete flying focus and driven plasma wave in the linear regime ($a_0=0.1$). (a) The on-axis $(|\mathbf{x}_\perp|=0)$ envelope of the DFF pulse (red and blue) and longitudinal electric field of the plasma wave (black and gray) have nearly identical profiles in $\xi = z-ct$ after the pulse has propagated 5 (solid) and 20 (dashed) dephasing lengths. (b,c) The spatiotemporal structure of the DFF envelope (top) and longitudinal field (bottom) are also nearly identical at these propagation distances. From left to right in $\xi$, the variation in the transverse width of the DFF (i.e., the X structure) results from $\xi$ slices that have already passed through focus, are at or near focus, and have yet to come into focus, respectively. This slice-dependent focusing is also the source of the pre and postcursors on the envelopes  in (a). In this example, $ N= 250 = N_\text{min} \gg N_\text{overlap}$. Due to significant overlap between individual pulses, the spot size of the overall envelope $k_p w_\text{eff} = 9$ is larger than that of the individual pulses $k_p w_0 = 4$.}
        \label{fig:Eenv-Ez-snapshots}
    \end{figure*}        
    
\section{Simulations}
\label{sec:simulations} 
    
    To demonstrate that a DFF can eliminate dephasing in LWFA, simulations were performed using the quasistatic, quasi-3D particle-in-cell (PIC) code QPAD \cite{li2021quasi}. QPAD models the evolution of the plasma electrons in response to the ponderomotive force of a laser pulse, as well as the electric and magnetic fields generated by the resulting charge separation. The induced electron current, in turn, feeds back onto the propagation of the pulse through the relativistic nonlinearities and ponderomotive modifications to the electron density profile. Further details regarding the algorithm and numerical parameters can be found in Appendix~\ref{sec:simulation-details}. Comparisons of QPAD to the full (i.e., non-quasistatic) PIC code OSIRIS \cite{Fonseca2002} are presented in Appendix~\ref{sec:simulation-compare}.

\subsection{Linear Regime}
    
    Figure~\ref{fig:Eenv-Ez-snapshots} shows that dephasing can be circumvented in the linear regime by using a DFF and provides insight into the DFF and plasma wave structure. A DFF with $N=250$ and $a_0 = 0.1$ drives a plasma wave with $v_w = c$ over 20 dephasing lengths, $L = 20 L_d$ (see Table~\ref{tab:ez-onaxis-params} for other parameters). The absence of dephasing and the velocity $v_w = c$ are evident in Fig.~\ref{fig:Eenv-Ez-snapshots}(a) by the nearly identical structures and locations of the pulse and plasma wave in $\xi$, i.e., neither recedes in $\xi$ over 20 dephasing lengths of propagation. This indicates that a highly relativistic electron ($v_z \approx c$) initially located in the maximum accelerating field will remain in that field over the entire length of the accelerator. 

    Figures~\ref{fig:Eenv-Ez-snapshots}(b) and (c) display the transverse structure of the DFF pulse and plasma wave. Both the DFF pulse and plasma wave maintain near-constant transverse profiles throughout propagation, which is critical to ensuring that an ultra-relativistic electron experiences a constant focusing force as it accelerates. The coherent superposition of the $N = N_\mathrm{min} \gg N_\mathrm{overlap}$ pulses produces a continuous DFF envelope with an ``X''-shaped structure, arising from the slice-dependent focal length. The temporal precursor of the DFF pulse, visible in Figs.~\ref{fig:Eenv-Ez-snapshots}(a) and (b), gradually recedes as each pulse comes into focus and is replaced by the ``postcursor" visible in Figs.~\ref{fig:Eenv-Ez-snapshots}(a) and (c).

\begin{figure*}[t]
        \centering
        \includegraphics[width=\linewidth]{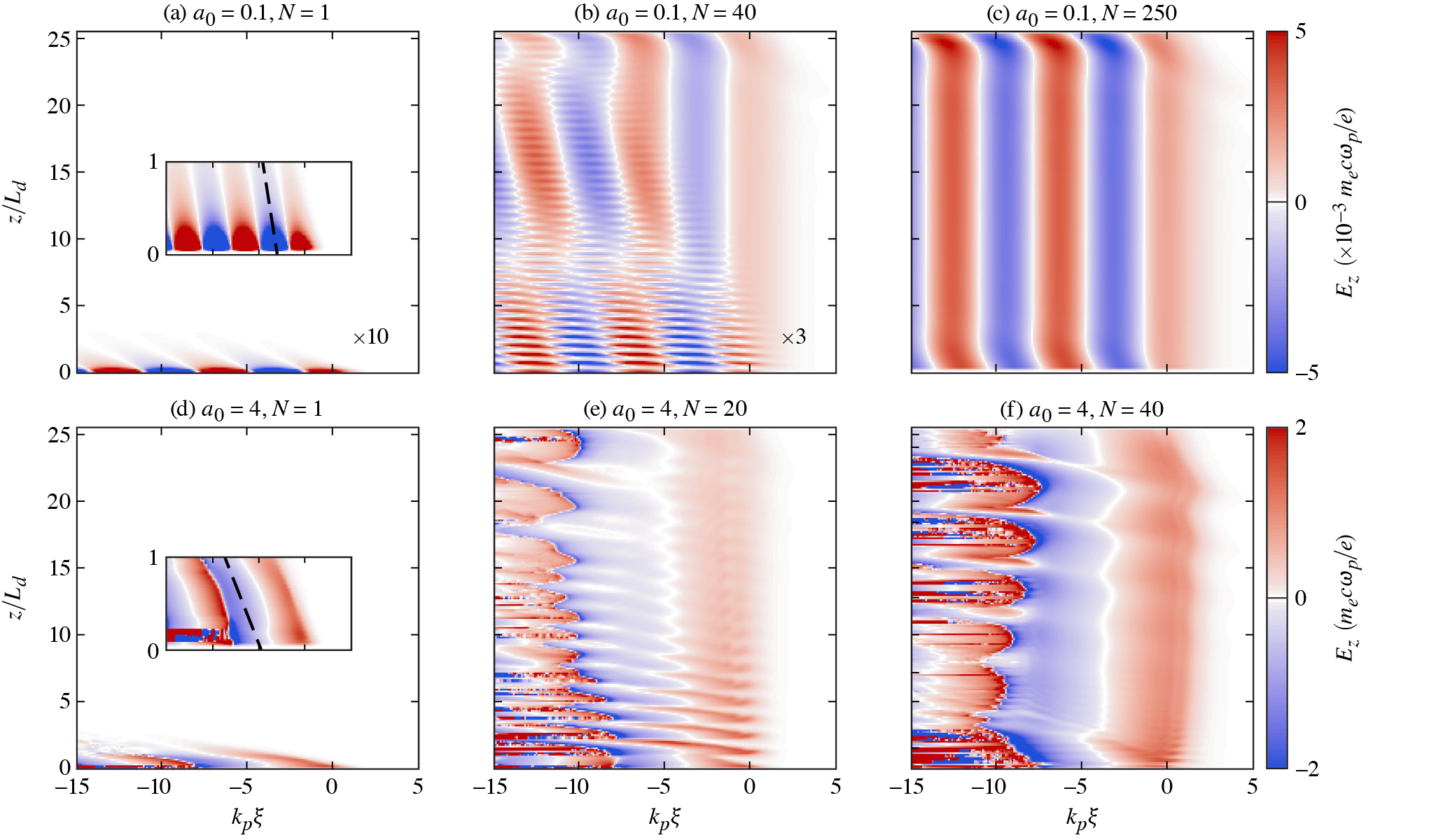}
        \caption{Dependence of the on-axis $(|\mathbf{x}_\perp|=0)$ longitudinal field $E_{z}$ on the number of pulses driving the plasma wave in the (a,b,c) linear and (d,e,f) nonlinear regimes. (a) With a single pulse ($N=1$) in the linear regime $(a_0 = 0.1)$, the maximum accelerating field (dashed black line) recedes and drops in amplitude due to the subluminal group velocity and diffraction of the pulse. (b) For a DFF composed of $N=40 < N_\text{min} = 250$ pulses, the plasma wave persists over the entire propagation range of 25 dephasing lengths, but the large separation between pulses results in an irregular longitudinal field. (c) A DFF with $N=250 =N_\text{min}$ pulses drives a plasma wave with a smooth longitudinal field and a phase velocity $v_w = c$. (d) With a single pulse ($N=1$) in the nonlinear regime $(a_0 = 4)$, the maximum accelerating field (dashed black line) recedes and drops in amplitude due to a combination of the subluminal group velocity and depletion. (e,f) Similar to the linear regime, a DFF composed of too few pulses ($N=20 < N_\text{min} = 50$) produces an irregular field, while a DFF composed of a sufficient number of pulses ($N=40 \approx N_\text{min}$) produces a smooth field.
        }
        \label{fig:ez-onaxis}
    \end{figure*}

Figures~\ref{fig:ez-onaxis}(a)--(c) show how the structure of the accelerating field ($E_z$) depends on the number of pulses composing a DFF in the linear regime. With a single pulse [Fig.~\ref{fig:ez-onaxis}(a)], the on-axis $(|\mathbf{x}_\perp|=0)$ accelerating field slides backward in $\xi$ and quickly drops in magnitude. These are direct results of the pulse driving a plasma wave with $v_w = v_g < c$ and quickly diffracting, respectively. The dashed line in the inset marks a trajectory traveling at $v_g$ over a single dephasing length. Thus, even if diffraction could be counteracted (for instance with a plasma waveguide), dephasing would still occur over a distance consistent with Eq.~\eqref{eq:Ld} ($k_p L_d \approx 2000$).

When the number of pulses is increased to $N = 40 < N_\mathrm{min}$, the accelerating field persists over the entire propagation distance of $25 L_d$ [Fig.~\ref{fig:ez-onaxis}(b)]. However, the phase velocity of the plasma wave $v_w$ evolves irregularly, and the longitudinal field exhibits discrete gaps due to the large separation between the focal points of each pulse. In this case, each pulse contributes to the drive over a long enough distance that spatiotemporal couplings develop in its profile. This invalidates the separable form of Eq.~\eqref{eq:singlepulse} (see Appendix \ref{sec:higher-order-effects}) and contributes a temporal broadening that smooths the longitudinal field at larger distances in Fig.~\ref{fig:ez-onaxis}(b). Despite this smoothing, a constant accelerating field over the entire acceleration distance still requires $N \gtrsim N_\mathrm{min}$.

Figure~\ref{fig:ez-onaxis}(c) demonstrates that a DFF with $N=N_\mathrm{min}=250$ pulses in the linear regime can excite a smooth and constant accelerating field with $v_w = c$. The parameters are the same as those in Fig.~\ref{fig:Eenv-Ez-snapshots}. A highly relativistic electron ($v_z \approx c$) initially collocated with the maximum accelerating field [blue stripe at $k_p\xi\approx-3$ in Fig.~\ref{fig:ez-onaxis}(c)] would travel along a vertical line in $\xi-z$ space and experience the maximum accelerating field over the entire 25 dephasing lengths.

\subsection{Nonlinear Regime}

While the linear regime provides important insights into the benefit of LWFA driven by a DFF, the nonlinear regime ($a_0\gtrsim2$) offers several key advantages. First, the accelerating field is much higher, allowing for significantly larger energy gains in a single stage. Second, the focusing fields are much stronger and linear in radius, which helps preserve the emittance. Finally, the DFF can be composed of far fewer pulses, which would reduce the complexity of an experimental implementation. This latter advantage arises because self-guiding allows each pulse in the DFF to strongly drive the plasma wave over its depletion length, rather than being limited to the Rayleigh range as in the linear regime.

Figures~\ref{fig:ez-onaxis}(d)--(f) illustrate how the structure of the on-axis accelerating field depends on the number of pulses composing the DFF in the nonlinear regime. With a single pulse [Fig.~\ref{fig:ez-onaxis}(d)], the accelerating field slides backwards in $\xi$, as it did in the linear regime. However, in contrast to the linear regime, the maximum amplitude of the accelerating field is relatively constant over a single dephasing length (see inset). This is because the individual pulse parameters were chosen to be nonlinearly matched for self-guiding ($k_p w_0 = 2 \sqrt{a_0}$), which prevents the drop in amplitude due to linear diffraction. Instead the amplitude drops over a depletion length $L_\mathrm{pd} =  L_d/2 \approx 10z_R$ as the pulse transfers its energy to the plasma wave. 

The dashed line in the inset of Fig.~\ref{fig:ez-onaxis}(d) marks the trajectory of the accelerating field driven by a single pulse in the nonlinear regime. The accelerating field slips backward $2.5\times$ more rapidly than in the linear regime [see Fig.~\ref{fig:ez-onaxis}(a) inset]. This is consistent with Ref.~\cite{lu2007generating}, which predicts a $3\times$ higher rate of slippage due to etching, i.e., local depletion at the front of the pulse. Although the rate of slippage is higher, the nonlinear dephasing length is similar to the linear dephasing length ($k_p L_d \approx 2000$) because the wavelength of the plasma wave is longer in the nonlinear regime. 

As in the linear regime, increasing the number of pulses in the nonlinear regime provides a smoother and more-constant accelerating field that persists over a longer distance [Figs.~\ref{fig:ez-onaxis}(e) and (f)]. However, the evolution of the accelerating field is now more complex. For $N = 20 < N_\text{min}$, the large separation between the pulses results in a longitudinal field with a modulated period [Fig.~\ref{fig:ez-onaxis}(e)]. In the nonlinear regime, the period of the plasma wave depends on the amplitude and spot size of the driving pulse, both of which oscillate as each pulse comes into focus and then depletes. At larger $z$, the modulations are partially smoothed due to the development of space-time couplings in the profile of each pulse, which increases their overlap.

Figure~\ref{fig:ez-onaxis}(f) demonstrates that a DFF with $N=40 \approx N_\mathrm{min}$ pulses in the nonlinear regime can excite a smooth and constant accelerating field with $v_w = c$. Because the overall pulse is more uniform, the modulations in the accelerating field observed with $N=20$ are mostly absent. In principle, the residual modulations could be mitigated by optimizing the parameters of each pulse, such as the relative delays, focal locations, durations, or spot sizes. This will be a topic for future investigation. For the examples in Figs.~\ref{fig:ez-onaxis}(e) and (f), the polarization alternated between the $\mathbf{\hat{x}}$ and $\mathbf{\hat{y}}$ directions from one pulse to the next. This reduces the interference between pulses and was found to produce a more stable accelerating structure.

With regards to the striations observable at smaller $\xi$ in Figs. 3(d)--(f), QPAD is accurate within the first period of the plasma wave, but can deviate from full OSIRIS simulations in the nonlinear regime after the first period (see Appendix C for comparisons). This is a consequence of the quasistatic approximation, which assumes that the background electrons have a velocity less than $v_w$. These deviations, however, do not impact the accelerating field experienced by a separate population of highly relativistic electrons located in the first period.

    \begin{figure*}[t]
        \centering
        \includegraphics[width=\linewidth]{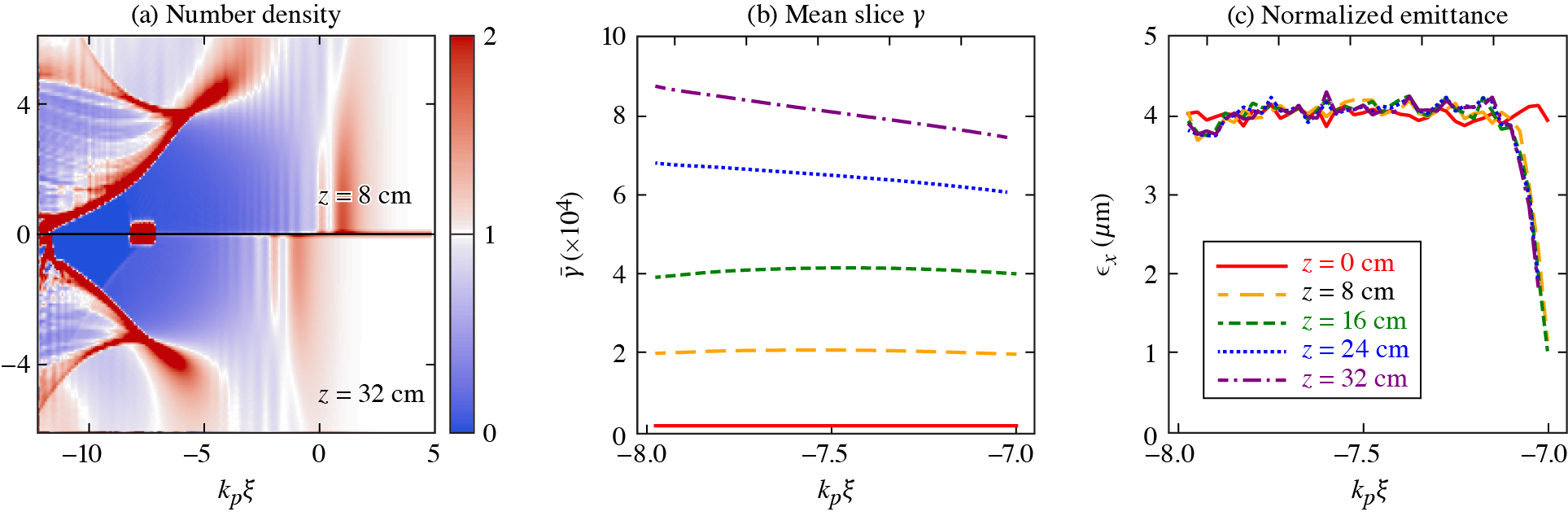}
        \caption{Acceleration of a $50 \; \mathrm{pC}$ electron beam from $1 \; \mathrm{ GeV}$ to $41 \; \mathrm{GeV}$ over $32 \;\mathrm{cm}$, or 50 dephasing lengths, using a discrete flying focus with tailored delays. (a) Electron density of the background plasma and electron beam normalized to $n_0$. The beam profile and accelerating structure are preserved over the entire length of the accelerator. Note that the beam density saturates the color map. (b) Mean slice energy of the beam. At the end of the accelerator ($z=32 \; \mathrm{cm}$), the spectrum is nearly monoenergetic with an energy spread $\sigma_\gamma/\bar{\gamma} = 0.04$. (c) Normalized slice emittance. The emittance does not grow as the beam is accelerated.}
        \label{fig:beam-evolution}
    \end{figure*}

    \subsection{Acceleration in the Nonlinear Regime}
    
    The plasma wave driven by a DFF in the nonlinear regime can accelerate an electron beam over many dephasing lengths while preserving the initial emittance of the beam. To demonstrate this, QPAD simulations were conducted with an axially uniform {48-pC}, 1-GeV monoenergetic electron beam externally injected into a plasma wave driven by a DFF with $a_0 = 4$, $N = 80$, and $L = 50 L_d = 32 \; \mathrm{cm}$. These are the same DFF parameters used for the example in Sec. II and for Fig. 3(f), but with twice the length. The length $L = 50 L_d$ was chosen to accommodate demands on computational resources and does not represent a physical limitation of DFF-driven LWFA. 

    Figure~\ref{fig:beam-evolution} shows that the DFF-driven plasma wave [Fig.~\ref{fig:beam-evolution}(a)] accelerates the injected electrons from 1 GeV to an average energy of $m_ec^2\bar{\gamma}= 41$ GeV [Fig.~\ref{fig:beam-evolution}(b)]. This is 3$\times$ higher than the energy achievable with a conventional pulse in a lower-density stage of the same length. Over the entire beam, the acceleration preserves the normalized slice emittance at its initial value of $\epsilon_x = 4 \; \mu$m [Fig.~\ref{fig:beam-evolution}(c)]. The average slice energy develops a slight positive chirp, which contributes to a final energy spread of $\sigma_\gamma/\bar{\gamma} = 0.04$. The slice energy spread, however, is significantly lower: $\smash{\overline{\sigma_\gamma(\xi)/\bar{\gamma}(\xi)}} = 0.001$. The slight decrease in the spot size of the beam, observable in Fig.~\ref{fig:beam-evolution}(a), is consistent with the theoretically predicted adiabatic evolution $\sigma_x(\gamma) \propto \gamma^{-1/4}$ \cite{khachatryan2007femtosecond}.
    
    For this example, the parameters of each pulse composing the DFF were tailored to prevent the injected beam from interacting with the rear sheath of the plasma wave. The delays $\Delta_j$ of the first two pulses were shifted back relative to Eq.~\eqref{eq:delta-j-linear} by $2k_p^{-1}$ and $1 k_p^{-1}$, respectively. All subsequent pulses had delays given by $ \Delta_j = \left[ \frac 1 2 (k_p / k_0)^2 - S \right] f_j $, where $f_j = 1250(j-1) k_p^{-1} \approx (5/4)(j-1)L_\mathrm{pd}$. Here, $S = 1.25\times10^{-5}$ is a correction to the linear group velocity factor $1-v_g/c \approx \frac 1 2 (k_p / k_0)^2 = 8\times10^{-4}$ that corresponds to an additional slippage of $1 \; k_p^{-1}$ over $40 L_d$ of propagation. In a separate QPAD simulation where uniform spacing was used for the $\Delta_j$ (as in Fig.~\ref{fig:ez-onaxis}), most of the beam was lost: the rear sheath of the plasma wave repeatedly swept over and defocused the beam.

    The initial normalized emittance of the beam, $\epsilon_x = 4 \; \mu$m, was chosen to satisfy the emittance matching condition for the initial spot size $k_p \sigma_x = 0.2$. This emittance value is relatively high in the context of linear collider designs. While the matched emittance can be decreased by using a smaller spot, the spot was limited in the simulations by computational demands on the transverse resolution $k_p \Delta x = 0.1$. To examine the emittance growth of a lower-emittance beam, QPAD simulations were also run with $L = 5 L_d$ (i.e., $10\times$ shorter distance), a $10\times $ higher transverse resolution of $k_p \Delta x = 0.01$, a spot size of $\sigma_x = 0.02 k_p^{-1} = 65 \text{ nm}$, and a matched normalized emittance of $\smash{\epsilon_{x} = 0.0126 k_p^{-1} = 40 \; \mathrm{nm}} $. These simulations also exhibited negligible emittance growth. This suggests that emittance growth would be negligible over a distance $L = 50 L_d $ if the beam were initialized with a collider-quality emittance.

    The DFF and electron beam parameters were chosen to demonstrate the concept and were not optimized for accelerated charge or efficiency. Nevertheless, it is still instructive to calculate the efficiency. The initial energy in the overall pulse was $150 \; \mathrm{J}$, while the total energy gain of the electron beam was $1.9 \; \mathrm{J}$, yielding an efficiency of $1.3\%$. This efficiency is low for two reasons. First, the beam needs to be in front of the rear sheath of the plasma wave because it oscillates, as seen in Fig.~\ref{fig:ez-onaxis}(f). Second, the blowout is incomplete, i.e., the background electrons have not been completely expelled in the first period of the plasma wave. If these issues can be resolved through optimization of the pulse train, the efficiency could be significantly higher. Equation 6 of Ref.~\cite{tzoufras2008beam} indicates that a total charge of $Q \approx 400  \; \mathrm{pC}$ with an optimized trapezoidal beam profile can be loaded in a fully blown out wave at $E_z = -m_e c \omega_p/e$, which would increase the efficiency to $ 10\%$. 

\section{Discussion and Conclusions}
\label{sec:discussion}

    A discrete flying focus can eliminate dephasing by driving a plasma wave with a phase velocity equal to the vacuum speed of light ($v_w = c)$. This allows highly relativistic electrons to remain collocated with the maximum accelerating field of the plasma wave over a distance limited only by the number of pulses or total pulse energy. By removing the constraint of dephasing, a DFF-driven LWFA can operate at higher densities, providing larger energy gains in a single stage or equal energy gains in a shorter stage. The former would reduce the number of stages required to reach a target electron beam energy.

    Quasi-3D PIC simulations with non-optimized parameters demonstrated that a DFF-driven plasma wave can accelerate a 50-pC electron beam from 1 GeV to 41 GeV over 50 dephasing lengths while preserving the beam quality. This is $3\times$ the maximum energy gain achievable with a conventional pulse in a lower-density stage of the same length. The acceleration length in the simulation was limited by computational resources and could, in principle, be extended to attain beam energies beyond {100 GeV} in a single stage. With a longer stage or at higher density, the energy gain achievable by a DFF becomes even larger than that of a conventional pulse [Eq.\eqref{eq:egainRatio}].

    The simulation domain was filled with pre-ionized plasma. In practice, this would require a large amount of energy, which would reduce the overall efficiency. For the 41-GeV result, the energy required to pre-ionize the hydrogen gas would be $ \pi R^2 L_p n_0 U_i \approx 2600 \; \mathrm{ J}$, where $R$ is the maximum radius of the simulation and $U_i = 13.6 \; \mathrm{eV}$ is the ionization potential. In principle, the energy required could be reduced by a factor of 3 if the ionization were structured into a cone. The energy required for ionization could also be substantially reduced if either (i) the DFF was used to ionize the gas in flight or (ii) the DFF could be efficiently coupled into a pre-ionized plasma with a radius of a few spot sizes $w_0$. In either case, refraction could be significant. This could potentially be addressed by tailoring the $\Delta_j$ sequence (Eq.~\eqref{eq:delta-j-linear}) and other properties to account for the differences from propagation in a uniform plasma. This is an area for future work.
    
    The DFF was designed using the formalism of ASTRL pulses \cite{pierce2023arbitrarily}. The configuration considered here consists of multiple, collinear pulses with tunable delays, focal points, and polarizations. The ability to independently tune these parameters provides a flexible and complimentary approach to continuous flying focus concepts and could facilitate optimization of the plasma formation, energy gain, or beam quality---albeit at the cost of increased complexity due to the use of multiple pulses. For instance, the techniques described in Refs. \cite{su2023optimization,nunes2024bayesian} could be adapted to determine a pulse train that mitigates refraction, produces a more complete blowout, or suppresses the oscillations in the rear sheath of the plasma wave observed in Figs.~\ref{fig:ez-onaxis}(e) and (f). More generally, the duration, axial shape, spot size, or orbital angular momentum of each pulse could also be tuned using the ASTRL formalism, offering extensive possibilities. A train of incoherent pulses could also be used to relax requirements on the coherence of each pulse. The alternating polarization case, which resembles an incoherent sum, provides confidence that this approach could be effective. 

    Here, the DFF was used solely to produce a plasma wave with $v_w = c$, but the flexibility to structure the pulse train could be beneficial in a variety of LWFA scenarios. For instance, a DFF could drive a plasma wave with a slower phase velocity for muon acceleration \cite{geng2024efficient}, or a superluminal phase velocity to prevent wave breaking \cite{palastro2021wavebreaking}. As another example, a DFF with orbital angular momentum---constructed by choosing a different Laguerre-Gaussian mode for the $C_j$---could be used for dephasingless LWFA of positrons  \cite{vieira2014nonlinear,palastro2020dephasingless}. Finally, the first few pulses could be tailored to either facilitate self-injection or to suppress it when external injection is employed. 

    An experimental realization of a DFF pulse could take advantage of techniques for manipulating and assembling multiplexed laser pulses. For instance, the multiplexed pulses of a high-power fiber laser \cite{Rainville2024} could be independently delayed and focused before being coherently combined on a collinear path. Alternatively, if the central frequency of each pulse can vary, a broad bandwidth pulse could be dispersed into several pulses that could be independently delayed and focused. Note, however, that dividing the bandwidth in this manner would place a lower bound on the duration $\tau_j$ of each pulse. The possibility of using different central frequencies and assessing the limitations thereof requires further investigation. A schematic and further discussion of possible techniques can be found in the original work on ASTRL, Ref. \cite{pierce2023arbitrarily}.

\begin{acknowledgments}

    The authors acknowledge Thamine Dalichaouch for helpful discussions on beam loading in LWFA. This work was supported by the University of Rochester, Laboratory for Laser Energetics subcontract SUB00000211/GR531765, the Department of Energy National Nuclear Security Administration under Award No. DE-NA0004131, and Department of Energy Awards No. DE-SC0025612 and  DE-SC0021057. This research used resources of the National Energy Research Scientific Computing Center (NERSC), a Department of Energy Office of Science User Facility using NERSC award HEP-ERCAP31590 (project mp113).

\end{acknowledgments}

\appendix 

\section{Higher-Order Effects of Linear Laser Propagation}
\label{sec:higher-order-effects}

    The laser profiles in Eqs.~\ref{eq:singlepulse} and \ref{eq:astrl} were derived by dropping three terms from Eq.~\eqref{eq:env-pde-full}: the mixed derivative $\partial_s \partial_\zeta$, the dispersive term $(1 - v_g/c)^2\partial^2_\zeta$, and the $\partial_s^2$ term. Each of these terms has a corresponding length scale $L_{s\zeta}$, $L_{\zeta\zeta}$, $L_{ss}$ that determines when higher-order corrections to the approximate solution in Eq.~\eqref{eq:singlepulse} may be important. These length scales can be estimated for a single pulse following the multiscale expansion approach in Ref.~\cite{pierce2023arbitrarily}:
    \begin{align}
        L_{s\zeta} &= \frac 1 2 c \tau_0 k_0^2 w_0^2 
        \label{eq:Lszeta}
        \\ L_{\zeta\zeta} &= \frac 1 2 \left( \frac{k_0}{k_p} \right)^2 k_0 (c \tau_0)^2
        \label{eq:Lzetazeta}
        \\ L_{ss} &= \frac 1 2 k_0^3 w_0^4
        \label{eq:Lss}
    \end{align}
   The $L_{s\zeta}$ and $L_{ss}$ length scales are the same in vacuum and plasma. For a pulse initialized according to Eq.~\eqref{eq:singlepulse}, $L_{s\zeta}$ is the length scale over which the envelope will develop non-separable dependencies on the temporal and transverse coordinates. Beyond the $L_{\zeta\zeta}$ length scale, group velocity dispersion causes substantial temporal broadening of the pulse, which reduces its amplitude. 
   Finally, when $L_{ss}$ is comparable to the Rayleigh range $z_R = \tfrac 1 2 k_0 w_0^2$, non-paraxial corrections will cause the transverse profile to deviate from a solution to the paraxial wave equation.    

    These estimates apply to a single pulse. Due to interference, the evolution of the pulse train is more complex. Nonetheless, the length scales for a single pulse provide estimates for when higher-order linear effects can contribute. When they do contribute, it may be necessary to tailor the pulse train to compensate. For example, the $L_{s\zeta}$ term results in a non-ideal radial group delay, where the amount of slippage depends on radius. This can be compensated by adjusting the $\Delta_j$ for each pulse. The correction can be estimated from geometric optics as $\smash{\Delta_j = \Delta_j^{(0)} + \Delta_j^{(1)}}$, where $\smash{\Delta_j^{(0)}}$ is the right-hand side of Eq.~\eqref{eq:delta-j-linear} and the first-order correction is given by 
    \begin{equation}
        \Delta_j^{(1)} = \frac{f_j}{k_0^2 w_0^2}.
        \label{eq:Lszeta-compensation}
    \end{equation}
    
    The simulation results in Section~\ref{sec:simulations} can be interpreted in light of these considerations. The single-pulse length scales using the parameters in Table~\ref{tab:ez-onaxis-params} are $k_p L_{s\zeta} = 7500$, $k_p L_{\zeta\zeta} = 2\times10^4$, and $L_{ss} = \num{10^4}z_R$.
    The correction in Eq.~\eqref{eq:Lszeta-compensation} is consistent with the evolution of the phase velocity in the case with $a_0 = 0.1$ and $N=40$ shown in Fig.~\ref{fig:ez-onaxis}(b), where there is a large separation between the individual pulses. For the case with $a_0 = 0.1$ and $N=250$ shown in Fig.~\ref{fig:ez-onaxis}(c), the phase fronts do not slip backward even though $L > L_{s\zeta}$ and $L > L_{\zeta\zeta}$. In this case, there is significant overlap between the pulses. The resulting interference produces an overall pulse with an effective spot size $k_p w_\text{eff} = 9 $ and duration $\omega_p \tau_\text{eff} = 4$ that are larger than those of the individual pulses. Using these values in Eqs. \eqref{eq:Lszeta} and \eqref{eq:Lzetazeta}, the length scales become $k_p L_{s\zeta} = 10^5 $ and $k_p L_{\zeta \zeta} = 1.25\times10^5$ so that higher-order effects are negligible over the simulated length scale $k_pL = 5\times10^4$. The $a_0=4$ cases in Sec. \ref{sec:simulations} demonstrate that despite propagation beyond $L_{s\zeta} $ and $L_{\zeta\zeta}$ and some distortion to the profile in Eq.~\eqref{eq:astrl}, it is possible to obtain good results without additional compensation to the delays $\Delta_j$ given by Eq.~\eqref{eq:delta-j-linear}. However, additional tailoring could potentially improve the results.

\section{Simulation Details}
\label{sec:simulation-details}

    QPAD is a quasistatic PIC code \cite{li2021quasi,li2022integrating}. Quasistatic PIC codes take advantage of several approximations to reduce computational expense: (i) the laser pulse or particle beam driver does not evolve in the time it takes a slice of plasma particles to pass through it;  (ii) all relevant waves have phase velocities near the speed of light; and (iii) in the case of a laser pulse, laser-cycle averaged (or ``ponderomotive guiding center") equations of motion can be used for the background electrons \cite{mora1997kinetic}. With these approximations, the full set of Maxwell's equations are simplified to a set of Poisson-like equations for the potentials of the plasma wave and a nonlinear wave equation for the envelope of the laser pulse. The quasi-3D algorithm expands the fields, charge and current densities, and laser envelope into azimuthal modes. These quantities are calculated on grids in the coordinates $(r,\xi)$, and the expansion is truncated at a desired mode number. For the simulations presented in this work, only mode 0 was used. 
    
    In all cases considered in this study, a uniform plasma was simulated. The radius of the plasma $R_p$ was large enough to fit the spot size $w_\text{max}$ of the most out-of-focus laser pulse at $z=0$. For the simulations with $k_p L = \num{50000}$ in Figs.~\ref{fig:Eenv-Ez-snapshots} and \ref{fig:ez-onaxis}, the most out-of-focus spot size was $ w_\text{max} \approx (L/z_R)w_0 = 1000k_p^{-1}$. These simulations used $R_p = 3300k_p^{-1}$. For the simulation with an externally injected beam in Fig.~\ref{fig:beam-evolution}, the plasma radius was $R_p \approx 6600 k_p^{-1} \approx 2\; \mathrm{cm} $. The initial Lorentz factor of the beam was $\gamma_0 = 2000$ so that $ m_e c^2 \gamma_0  \approx  1 \; \mathrm{ GeV}$. The beam was cylindrically symmetric with a spot size $ \sigma_x = 0.2 k_p^{-1} = 650 \; \mathrm{nm}$ and a matched normalized emittance $\smash{\epsilon_{x} = k_p \sigma_{x}^2  (\gamma_0 / 2)^{1/2} = 1.26 k_p^{-1} = 4 \text{ }\mu\text{m}}$ \cite{khachatryan2007femtosecond}. The current profile was constant in $\xi$ with a length $k_p L_b = 1$. The resolution in all cases was $ k_p \Delta z = 0.01 $, $k_p \Delta x_\perp = 0.1$, and $ \omega_p \Delta t = 10$. These values were chosen to ensure numerical convergence in the cases where $a_0=4$. 
    
    \begin{figure}
        \centering
\includegraphics[width=\linewidth]{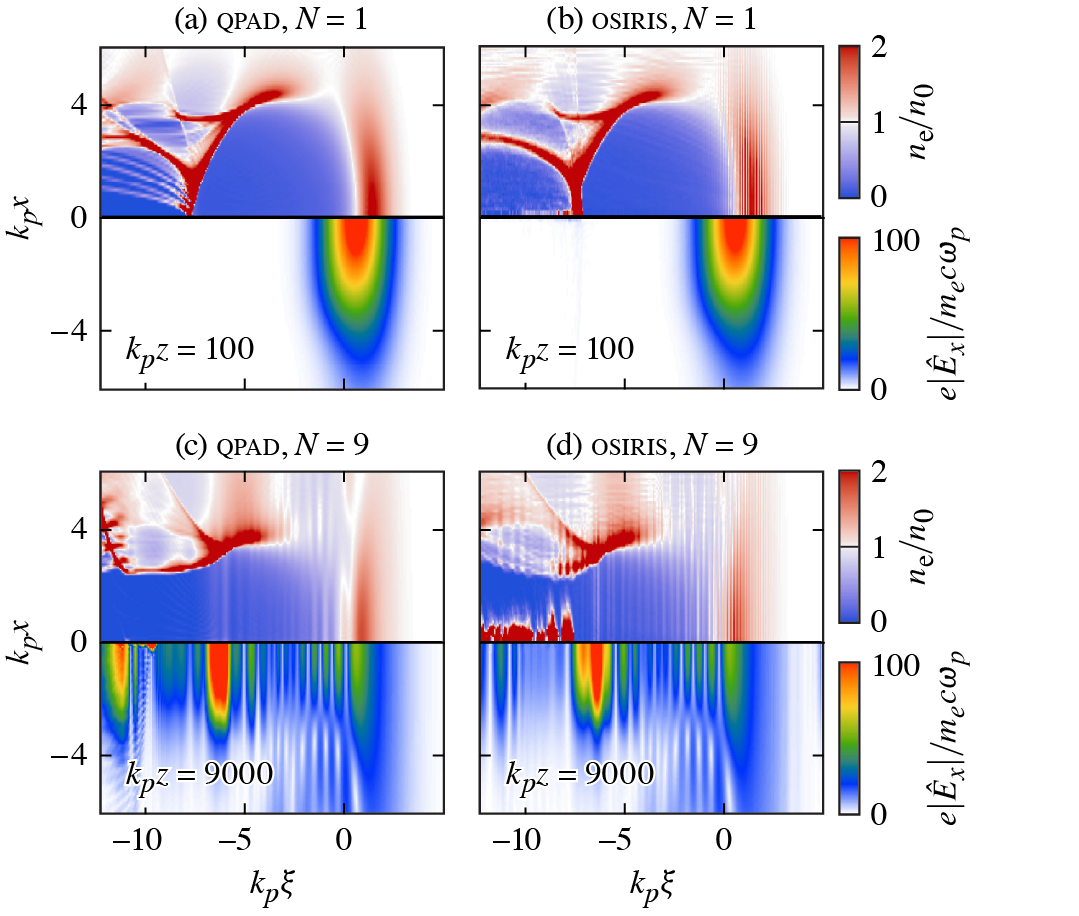}
        \caption{The electron densities and electric field envelopes from (left) QPAD and (right) OSIRIS simulations run with the same physical parameters. The results for a single pulse ($N=1$) are shown in (a) and (b), while the results for a DFF with $N=9$ pulses are shown in (c) and (d). The agreement between QPAD and OSIRIS is excellent. In (c) and (d), local depletion and redshifting causes the light to slip backward, where it is temporarily trapped by the plasma wave. This prevents the rear sheath from closing. Over longer distances, the sheath undergoes multiple cycles of opening and closing as the trapped light first accumulates and then escapes out the back. [See Figs.~\ref{fig:ez-onaxis}(e) and (f)]. The quasistatic model used in QPAD does not capture self-injection, which is observed in the OSIRIS simulations (d).}
        \label{fig:charge-laser-snapshots}
    \end{figure}

\section{Nonlinear LWFA in QPAD and OSIRIS}
\label{sec:simulation-compare}
    In order to demonstrate the accuracy of QPAD, reduced-scale simulations were performed to compare QPAD and the fully electromagnetic PIC code OSIRIS \cite{Fonseca2002} using identical laser pulse and plasma parameters. Remarkable agreement between the electron densities and the electric field envelopes of the laser pulses is evident in Fig.~\ref{fig:charge-laser-snapshots}. The physical parameters correspond to the nonlinear ($a_0 = 4$) cases presented in Table I, but with $N = 1$ [Figs.~\ref{fig:charge-laser-snapshots}(a) and (b)] or $N = 9$ [Figs.~\ref{fig:charge-laser-snapshots}(c) and (d)].

    Figures~\ref{fig:charge-laser-snapshots}(a) and (b) show the electron densities and electric field envelopes of the laser pulses from (a) QPAD and (b) OSIRIS for $N=1$ at $k_p z = 100  \ll k_p L_\text{pd}$. Within the first period of the plasma wave, the structure of the waves and envelopes agree. For $k_p\xi\lesssim -7$, the QPAD results deviate from the OSIRIS results: In the nonlinear regime, background electrons in the rear sheath of the wave can be accelerated to velocities close to $v_w$, which causes the quasistatic approximation to break down.
    
    Figures~\ref{fig:charge-laser-snapshots}(c) and (d) show the electron densities and electric field envelopes of DFF pulses from (c) QPAD and (d) OSIRIS for $N=9$ at $k_p z = \num{9000}$. A major difference from the $N=1$ case is the highly modulated and lengthened envelope of the pulses. As the individual pulses in the DFF deplete, they are locally redshifted. The lower group velocity of the redshifted light causes it to slide back into the plasma wave structure. This redshifted light can become trapped in the wave and modify its structure---for example, by preventing closure of the rear sheath, as seen in Figs.~\ref{fig:charge-laser-snapshots}(c) and (d). For longer stages, the redshifted light can cause multiple cycles of sheath opening and closing, as observed in Fig. \ref{fig:ez-onaxis}(f). The local wavenumber of the light (not shown) drops from $k(\xi) = k_0 = 25k_p $ at $\xi=0$ to $k(\xi) = 12 k_p\approx k_0/2 $ at $\xi\approx -6$. The longitudinal resolution $\Delta \xi = 0.01$ in the QPAD simulations was chosen to resolve this redshifting. 
    
    Another difference between the $N=1$ and $N=9$ cases is the more complete blowout in the $N=1$ case. The incomplete blowout driven by the DFF (on-axis density ${\approx}0.1 \; n_0$) reduces the accelerating field and limits how much charge can be loaded in the plasma wave. The incomplete blowout is caused by a reduction in the DFF amplitude, which falls to about half its original amplitude. The amplitude reduction could potentially be addressed by further tailoring of the laser pulses.
    
    The electron density from the $N=9$ OSIRIS simulation features a 0.35-nC self-injected bunch [Fig.~\ref{fig:charge-laser-snapshots}(d)]. Self-injection cannot be not modeled within the quasistatic approximation \cite{morshed2010}. Nevertheless, QPAD accurately models the laser pulse and plasma evolution in the region upstream of the self-injected bunch, where the dynamics are causally disconnected from the self-injected bunch. In this region, QPAD effectively cycle-averages the short-wavelength oscillations in the plasma density observed in the OSIRIS simulations. Self-injection only occurs in the vicinity of the moving density spike at the rear of the sheath. Thus, self-injection does not alter the evolution of the externally injected beam in Fig.~\ref{fig:beam-evolution}.
    However, the OSIRIS simulation suggests that the design presented in Fig.~\ref{fig:beam-evolution} could have dark current. This dark current could potentially be mitigated by beam loading or tailoring the properties of the individual DFF pulses to stabilize the rear sheath.  

    Figures~\ref{fig:charge-laser-snapshots}(c) and (d) also indicate that redshifted light would be spatially overlapped with an externally injected electron beam, as used in Fig~\ref{fig:beam-evolution}. The coupling between this light and the electron beam (i.e., direct laser acceleration \cite{Pukhov1999}) is not currently modeled in QPAD. In principle, direct laser acceleration could lead to an increase in the transverse momentum and emittance of the beam. In practice, the effect of direct laser acceleration on the beam is negligible: The beam betatron wavenumber $k_\beta$ and the wavenumber of the redshifted light $k_L$ are always far from the DLA resonance condition $k_\beta = k_L$. For the parameters used here, $k_L \approx 12k_p $ and $\smash{k_{\beta}(t) < k_\beta(0) = k_p( 2\gamma_0)^{-1/2}  = 0.016 k_p} $. To confirm that direct laser acceleration is negligible, the equations of motion for a subset of externally injected electrons were numerically solved using laser pulse and plasma wave fields derived from a phenomenological model, as in Ref. \cite{miller2023accurate}. These calculations indicated that the emittance growth from the beam--laser interaction was negligible for the case in Fig.~\ref{fig:beam-evolution} with $k_p L = \num{100000} $ when using a beam with a spot size $\sigma_{0x} = 0.02 = 65 \; \mathrm{nm}$ and matched emittance $ \epsilon_{x} = 0.0126k_p^{-1} = 40 \; \mathrm{nm}$. 

    In Fig.~\ref{fig:charge-laser-snapshots}, the electric field envelope for QPAD was calculated from the complex vector potential (amplitude and phase), while the electric field envelope for OSIRIS was extracted using a Hilbert transform. The OSIRIS simulations were run using the quasi-3D azimuthal decomposition \cite{Davidson2015} and the field solver introduced in Ref.~\cite{Li2021} to mitigate numerical dispersion.

\bibliography{bib}

\end{document}